\pdfoutput=1
\documentclass[apj]{emulateapj}
\usepackage{graphicx}
\usepackage{wrapfig}
\usepackage[all]{xy}
\usepackage[usenames]{color}

\newcommand{\nc}{\newcommand}
\nc{\beq}{\begin{equation}}
\nc{\eeq}{\end{equation}}
\nc{\bea}{\begin{eqnarray}}
\nc{\eea}{\end{eqnarray}}
\nc{\n}{\nonumber \\}

\nc{\cm}{21$\,$cm }

\begin{document}  

\title{Probing the Dark Ages at  z$\sim$20: The SCI-HI 21$\,$cm All-Sky Spectrum Experiment}

\author{Tabitha C. Voytek \altaffilmark{1,*}, Aravind Natarajan \altaffilmark{1,2}, Jos\'{e} Miguel J\'{a}uregui Garc\'{i}a \altaffilmark{3}, 
Jeffrey B. Peterson \altaffilmark{1}, Omar L\'{o}pez-Cruz \altaffilmark{3}}
\altaffiltext{*}{tcv@andrew.cmu.edu}
\altaffiltext{1}{McWilliams Center for Cosmology, Carnegie Mellon University, Department of Physics, 5000 Forbes Ave., Pittsburgh PA 15213, USA}
\altaffiltext{2}{Department of Physics and Astronomy \& Pittsburgh Particle physics, Astrophysics and Cosmology Center, University of Pittsburgh, 100 Allen Hall, 3941 O'Hara Street, Pittsburgh, PA 15260} 
\altaffiltext{3}{ Instituto Nacional de Astrof\'{i}sica, Optica y Electr\'{o}nica (INAOE), Coordinaci\'{o}n de Astrof\'{i}sica, Luis Enrique Erro No. 1 Sta. Ma. Tonantzintla, Puebla, 72840 Mexico}

\begin{abstract}
We present first results from the SCI-HI experiment, which we used to measure the all-sky-averaged \cm brightness temperature in the redshift range 14.8$<$$z$$<$22.7. The instrument consists of a single broadband sub-wavelength size antenna and a sampling system for real-time data processing and recording. Preliminary observations were completed in June 2013 at Isla Guadalupe, a Mexican biosphere reserve located in the Pacific Ocean. The data was cleaned to excise channels contaminated by radio frequency interference (RFI), and the system response was calibrated by comparing the measured brightness temperature to the Global Sky Model of the Galaxy and by independent measurement of Johnson noise from a calibration terminator. We present our results, discuss the cosmological implications, and describe plans for future work.
\end{abstract}

\maketitle

\section{Introduction}
Emission and absorption due to the \cm spin flip transition of neutral Hydrogen (HI) have emerged as valuable tools to probe the physics of the Universe across a set of cosmological eras ranging from the pre-star dark ages to the epoch of reionization to the era of acceleration \citep{Loeb_Zaldarriaga_prl_2004, cooray_prd_2004, bharadwaj_ali_mnras_2004, furlanetto_briggs_astron_2004_a, furlanetto_briggs_astron_2004_b, furlanetto_oh_briggs_physrep_2006, pritchard_loeb, pritchard_loeb_review, Liu_etal}. Experiments studying the universe through the \cm transition are of two kinds: mapping experiments that measure the power spectrum of \cm fluctations and all-sky absolute brightness experiments that measure the averaged \cm brightness temperature. Mapping experiments include PAPER \citep{paper1, paper2}, GMRT-EoR \citep{gmrt_paciga}, LOFAR \citep{lofar}, MWA \citep{mwa}, and the proposed \citet{hera} and \citet{ska}. All-sky experiments include EDGES \citep{edges}, \citet{leda} and the proposed Dark Ages Radio Explorer (DARE) \citep{dare}. 

The first stars to form are composed almost entirely of primordial elements. These stars, called Pop. III.1, are thought to form in dark matter minihalos of mass $\approx 10^6 - 10^8 M_\odot$ at a redshift $z \approx$ 20-30  \citep{Bromm_Coppi_Larson_1, Bromm_Coppi_Larson_2, Bromm_Larson, Abel_Bryan_Norman, Omukai_Palla1, Omukai_Palla2, Tan_McKee1, Tan_McKee2, Bromm}, and provide UV illumination of the HI in their surroundings. Supernovae from these short lived, massive stars enriched the Universe with heavy elements, paving the way for the subsequent generation of lower mass stars and planets. Formation of the first stars is thus a subject of broad impact, but it is difficult to constrain Pop. III.1 star formation models due to the absence of observations at high redshifts. The observations and instrument described here aim to provide such constraints. 

Formation of the first stars results in the production of Lyman-$\alpha$ photons, which are efficient in coupling the spin ($T_{\rm s}$) and kinetic ($T_{\rm k}$) temperatures of HI through the Wouthuysen-Field mechanism \citep{lya1,lya2}.  At $z \approx$ 20-30 the neutral gas kinetic temperature $T_{\rm k} \propto (1+z)^2$ is well below the CMB temperature ($T_{\gamma}$). The Wouthuysen-Field mechanism sets $T_{\rm s} = T_{\rm k}$, so the turn-on of the first stars should produce a significant decrease in the \cm brightness temperature ($T_{\rm b}$) around the redshift of first star formation, with $T_{\rm b} \propto {(T_{\rm s} - T_\gamma)} / T_{\rm s} \approx -T_\gamma / T_{\rm k}$.  Later, heating of the gas by high energy radiation sets $T_{\rm s} = T_{\rm k} \gg T_\gamma$.  $T_{\rm b}$ then enters the emission saturation regime where it is small and positive. These two events bracket a trough in the \cm all-sky brightness temperature. The magnitude and width of the trough provide valuable information on the properties of the first stars and X-ray sources. This trough is also a more easily measurable feature in the all-sky spectrum \citep{shaver,madau,tozzi,Liu_etal} than the slope in the spectrum that occurs at lower redshifts as the universe is reionized. 

In the following sections we describe the experiment $''$Sonda Cosmol\'{o}gica de las Islas para la Detecci\'{o}n de Hidr\'{o}geno Neutro$''$ (SCI-HI). This experiment measures the \cm brightness temperature using a single antenna and attendant sampling system operating on-site in a radio quiet location. Calibrated, Galaxy-subtracted data collected June 1-15, 2013 is presented and cosmological implications are discussed.

\section{Experimental Setup}
In order to probe the \cm all-sky spectrum, we designed a single antenna experiment that is optimized to work from 40-130 MHz, with over 90\% antenna coupling efficiency from 55-90 MHz. The entire system is designed to run off 12V DC batteries, allowing the experiment to be portable and easy to set up in remote locations. 

\subsection{HIbiscus Antenna}
Our antenna design started with the need to observe over an octave of frequency. Using FEKO, a finite element simulation, a four-square antenna scaled to 70 MHz was modified by dividing the square plates into inclined trapezoidal facets. Additional panels were added to the sides of each petal, forming a strip line with fixed gap between the petals. The exact size and angles of the facets and the strip line gap were tuned to improve performance. A 1:6 scale model was then built with the best design and was tested in an antenna range to determine the actual antenna bandwidth, impedance and beam shape.

\begin{figure}[!tbh]
\flushleft
\scalebox{0.33}{\includegraphics{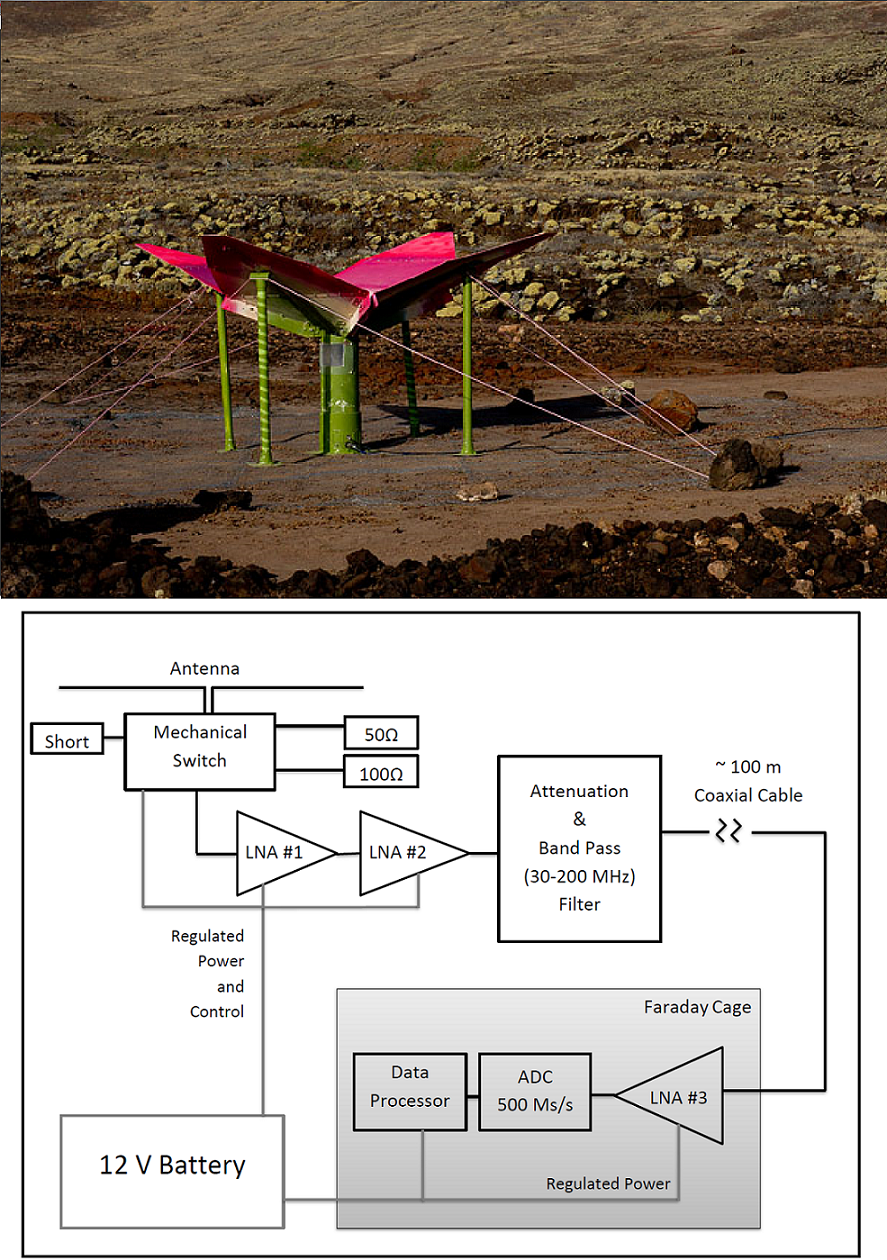}}
\caption{SCI-HI experiment. Top: HIbiscus Antenna on-site on Isla Guadalupe, Mexico. Bottom: System Block Diagram \label{antenna}}
\end{figure}

Critical parameters are the strip line gap and the height above the ground plane. The first parameter changes the impedance and the second parameter impacts the beam pattern. The antenna is fixed and points to the zenith. Our final design has increased bandwidth compared to the four-square antenna, a flattened impedance curve and a $55^{\circ}$ beam at 70 MHz with roughly frequency independent shape. The scale model measurements provide a check of the simulated antenna patterns that are used in the system calibration, as is shown in Eqn. \ref{eq:t_gsm}. A more detailed description can be found in (J\'{a}uregui-Garc\'{i}a et al in preparation) and the antenna is shown in Fig \ref{antenna}. We call this new design the HIbiscus antenna.

\subsection{Sampling system}
A basic block diagram of the instrument is shown in Fig. \ref{antenna}. The signal from our antenna passes through a series of electronic stages, including amplifiers and filters to remove radio frequency interference (RFI) below 30 MHz and aliasing of signals above 200 MHz. To measure the noise of the system, an electro-mechanical switch is placed between the antenna and the first amplifier. This switch, controlled by the sampling software, allows collection of spectra from terminators of known temperature ($50\Omega$, $100\Omega$ and Short). These spectra are collected in 5 min intervals and are used for calibration, as is shown in Eqns. \ref{eq:cal} and \ref{eq:cal1}.

The sampling system is composed of a GE PCIe digitizer board (ICS1650) for data acquisition and a PC with software for power spectrum generation. The signal from the antenna is sampled at 500 MSamples/sec with 12 bits of resolution, after which Fourier transform routines integrated into the sampling software are used to generate power spectra from 0-250 MHz. The system is placed inside of a Faraday cage $\sim$50 meters from the antenna and all power/control signals sent to the antenna and electronics as well as power to the internal PC buses are RF-filtered to minimize self-generated noise contamination. To minimize data collection interruptions, an external supervisor circuit monitors the system to provide restart in case of failure. The data sampling/processing duty cycle is $\sim$10\%, so 1 day of observation yields about 2 hours of effective integration time. 

\subsection{Experimental Sites}
One major noise contribution in the 40-130 MHz frequency band is terrestrial RFI. In particular, television stations and FM radio stations transmit within this band. Even at the U.S. National Radio Quiet Zone in Green Bank, West Virginia, the FM signal exceeds the sky signal by 10 dB over the entire FM band of 88-108 MHz. 

In order to bring RFI down below the level of the expected \cm signal, it is necessary to operate the instrument at an extremely remote radio-quiet location. For the data presented here, we travelled to Isla Guadalupe in Mexico. Isla Guadalupe (Latitude $28^{\circ}$ $58'$ $24''$ N, Longitude $118^{\circ}$ $18'$ $4''$ W) is 260 km off the Baja California peninsula in the Pacific Ocean. It is a Mexican biosphere reserve and has minimal infrastructure. As the location is a biosphere reserve, we were careful to ensure that our experiment had limited impact on the fragile insular ecosystem. 

We spent two weeks on Isla Guadalupe (June 1-15, 2013) collecting data on the western side of the island. Even at this remote site, we still detect some RFI from the mainland, although residual FM is only about 0.1 dB ($\leq$70 K) above the Galactic foreground level and there are no other strong RFI signals in our band.

\section{Results} 
\begin{figure}[!tbh]
\flushleft
\scalebox{0.50}{\includegraphics{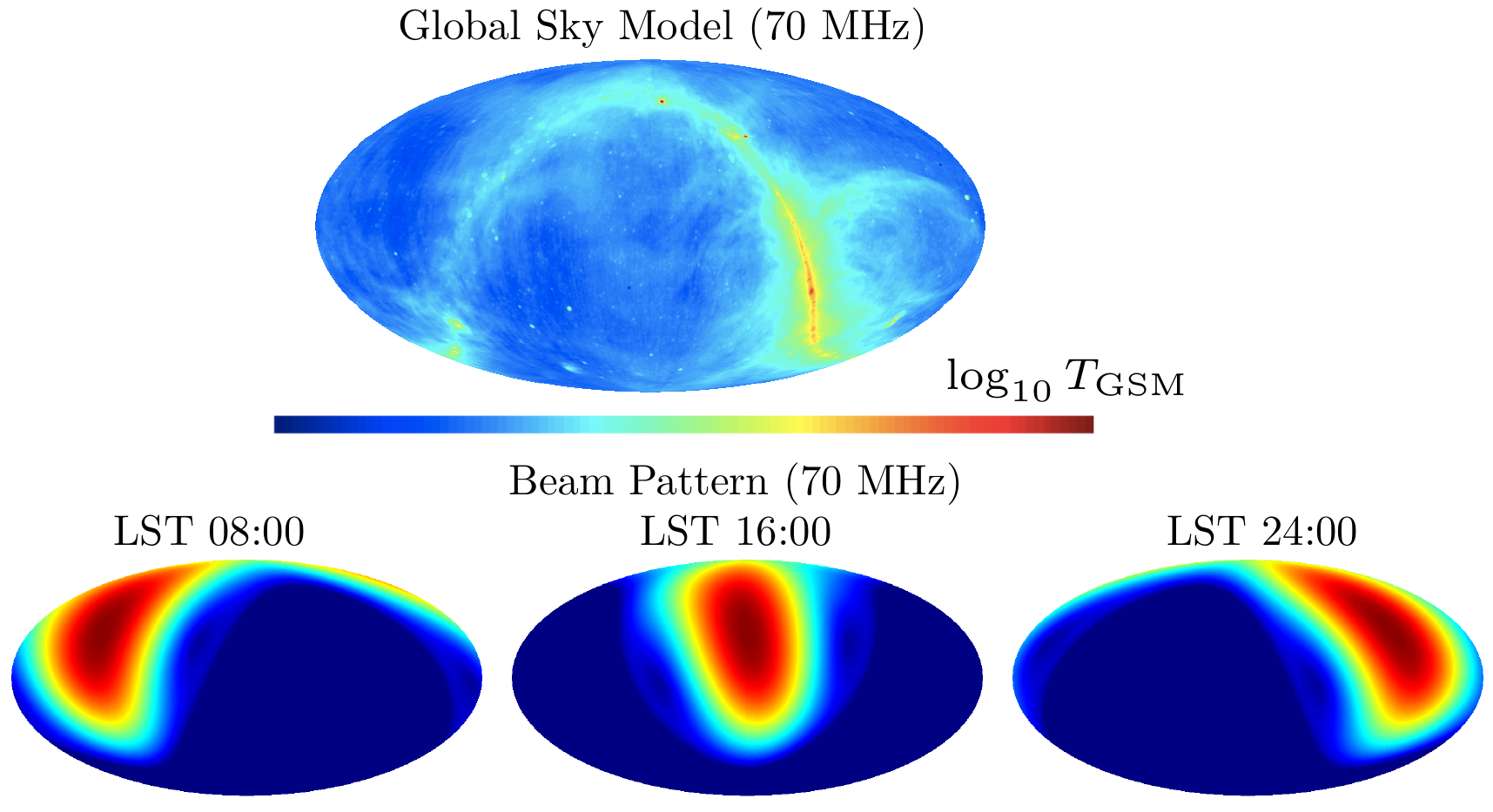}}
\caption{ Sky temperature and antenna beam pattern in (RA, DEC) coordinates. The top row shows the sky temperature (logarithmic) at 70 MHz, from the Galactic Global Sky Model (GSM). The bottom row shows the simulated antenna beam pattern at 70 MHz at different LST, plotted for the latitude of Isla Guadalupe.
\label{gsm_beam} }
\end{figure}

The Galactic Global Sky Model (GSM) \citep{gsm} provides a set of maps of the Galaxy at frequencies between 10 MHz and 100 GHz, created by interpolating data derived from publicly available large area radio surveys. GSM software is used to produce a set of maps of the sky in the frequency range relevant to our observations. For each frequency, the beam-averaged sky temperature as a function of time may then be obtained:
\beq \label{eq:t_gsm}
T_{\rm GSM}(t,\nu) =  \frac{ \int d\Omega \; {\rm GSM}(\theta,\phi,\nu) \; \mathcal{B}(\theta-\theta_0(t),\phi-\phi_0(t),\nu) } {\int \; d\Omega \; \mathcal{B}(\theta-\theta_0(t),\phi-\phi_0(t),\nu) }
\eeq
where $(\theta, \phi)$ are (RA, DEC) co-ordinates, $\mathcal{B}(\theta, \phi,\nu)$ is the simulated antenna beam pattern, and $(\theta_0(t), \phi_0(t))$ is the beam center. The time dependence of $(\theta_0, \phi_0)$ arises due to the rotation of the earth. Fig. \ref{gsm_beam} shows the GSM map at 70 MHz (top row) as well as the simulated antenna beam pattern for 3 different values of Local Sidereal Time (LST). The sky brightness is greatest when the beam center crosses the Galactic plane, and smallest when the Galactic plane aligns with the horizon. 

The spectra from our observations are cleaned to excise RFI contaminated data. Spectra are then calibrated using three techniques to determine the system gain $K (\nu)$, defined by the following equation:
\beq \label{eq:cal}
T_{\rm meas}(t,\nu) = K(\nu) \left [ \frac{P_{\rm meas} (t,\nu) }{\eta (\nu)} - P_{\rm short} (\nu) \right ]
\eeq
where $P_{\rm meas}(t,\nu)$ is the measured power spectrum, $\eta (\nu)$ is the transmission efficiency of the antenna, measured on-site immediately prior to data collection, and $P_{\rm short}(\nu)$ is the measured power spectrum from the short. 
 
The first technique uses the Johnson-noise calibration power spectra ($P_{\rm short} (\nu)$ and $P_{\rm 50 \Omega}(\nu)$) and the ambient temperature ($T_{\rm amb}$) to calculate the system gain $K_{\rm JNC} (\nu)$ directly using:
\beq \label{eq:cal1}
K_{\rm JNC} (\nu) = \frac{T_{\rm amb}}{P_{\rm 50 \Omega} (\nu) - P_{\rm short} (\nu)} 
\eeq

The second and third techniques both rely on the GSM estimate of the sky temperature and fit for $K (\nu)$ using a $\chi^2$ fit of the diurnal variation of the Galactic signal. For the second technique: 
\beq \label{eq:cal2}
\chi^2 (\nu) =  \sum_t \left [ T_{\rm meas}(t,\nu) - T_{\rm GSM}(t,\nu) \right ]^2
\eeq
is minimized to yield a best fit $K_{\rm GSM} (\nu)$.

For the third technique the daily mean signal is subtracted from both the $T_{\rm meas} (t,\nu)$ and the $T_{\rm GSM} (t,\nu)$ data prior to $\chi^2$ fitting: 
\beq \label{eq:cal3}
\chi^2 (\nu) = \sum_t \left [ \Delta T_{\rm meas}(t,\nu)- \Delta T_{\rm GSM} (t,\nu) \right ]^2
\eeq
where $\Delta T_{\rm meas}(t,\nu)= T_{\rm meas}(t,\nu)-\langle T_{\rm meas} \rangle_{\rm DAY}(\nu)$ and $\Delta T_{\rm GSM} (t,\nu) = T_{\rm GSM} (t,\nu) - \langle T_{\rm GSM} \rangle_{\rm DAY} (\nu)$. The minimization yields a best fit $K_{\rm \Delta GSM} (\nu)$. This third technique is analogous to an on-source/off-source calibration of a single dish radio telescope. 

\begin{figure}[!tbh]
\flushleft
\scalebox{0.43}{\includegraphics{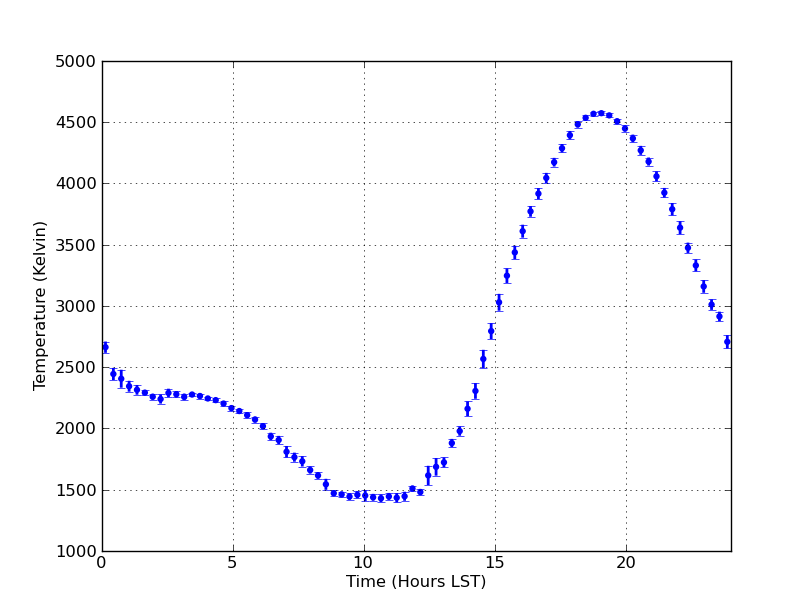}}
\caption{Diurnal variation of a single 2 MHz wide bin centered at 70 MHz. Calibrated mean with RMS error bars from day-to-day variation are shown for 9 days of observation binned in $\sim$ 18 minute intervals. Larger error bars correspond to LSTs where the quantity of useable data is smaller.  \label{diurnal}}
\end{figure}

Fig. \ref{diurnal} shows the daily time dependence of $T_{\rm meas}(t,\nu)$ at a single frequency. For some days, data was not collected for the entire 24 hour period and the mean signal varies with the covered LST range. 

\begin{figure}[!th]
\flushleft
\scalebox{0.62}{\includegraphics{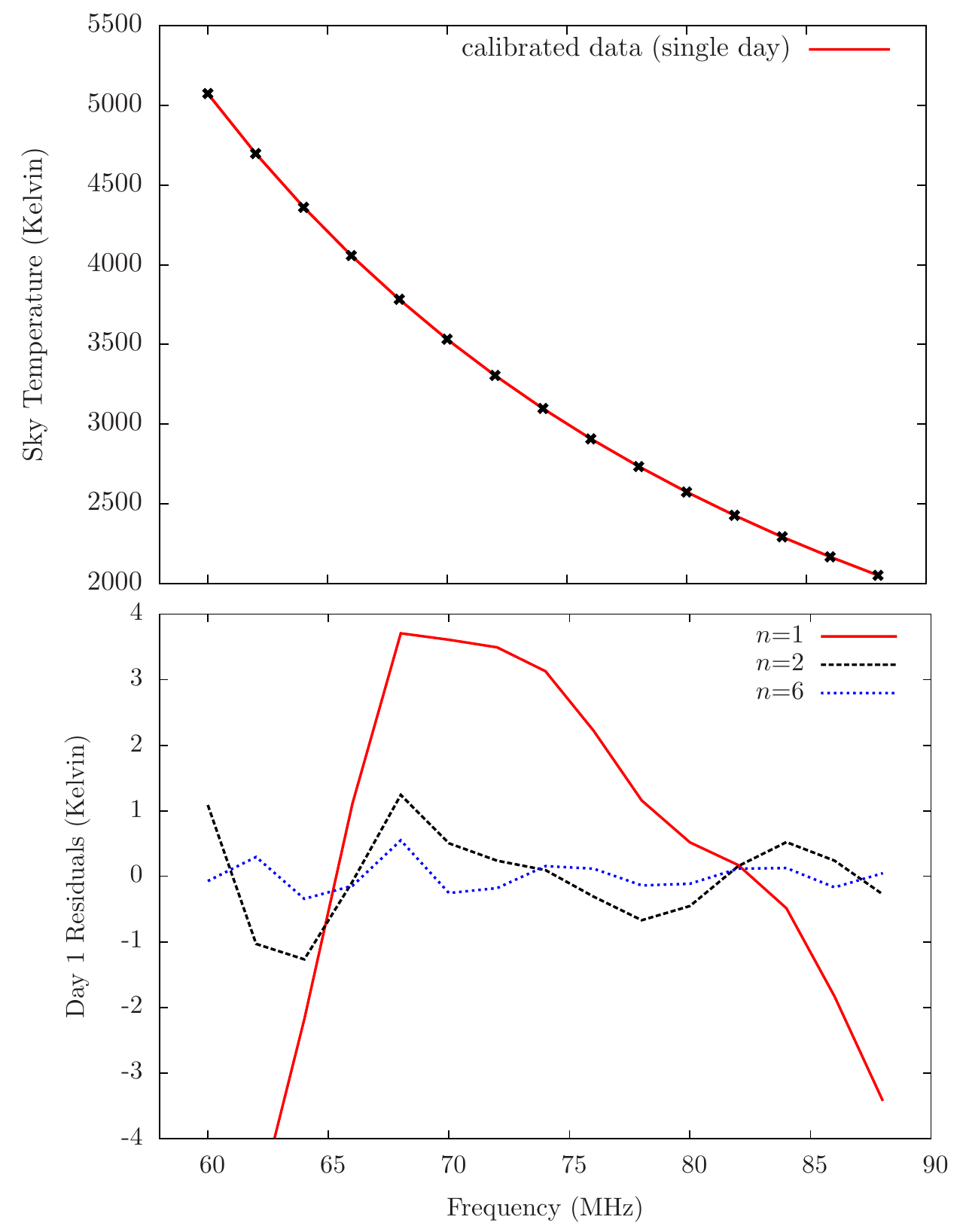}}
\caption{Data calibrated using $K_{\rm GSM}$ \citep{gsm}. The top plot shows mean data from a single day of observation ($\sim$50 minutes integration time) binned in intervals of 2 MHz. The best fit plot for $n$=2 is also shown ($a_0 = 3.548$, $a_1 = -2.360$, $a_2 = -0.164$, for June $1^{st}$). The bottom plot shows the residuals after the fit was subtracted. A simple power law ($n$=1) results in a poor fit, while $n$=2 substantially improves the fit. 
\label{raw_data} }
\end{figure}

After determining $K (\nu)$, each day of data is averaged to provide a single spectrum $\langle T_{\rm meas} \rangle_{\rm DAY} (\nu)$. This spectrum is fit to a model of the Galactic sky-averaged brightness temperature ($T_{\rm GM} (\nu)$):
\beq \label{eq:fit}
\log_{10} T_{\rm GM}(\nu) = \sum_{k=0}^n a_k \left [ \log_{10} \left( \frac{\nu}{70 \; {\rm MHz} } \right ) \right ]^k
\eeq
Using the calculated $a_k$ for each day of data, the residuals $\Delta T (\nu) = \langle T_{\rm meas} \rangle_{\rm DAY}(\nu)-T_{\rm GM} (\nu)$ are calculated. These $\Delta T (\nu)$ values are our estimate of the \cm all-sky brightness temperature spectrum after removal of Galactic emission. As shown in Fig \ref{raw_data}, there is substantial improvement in the fit obtained by increasing $n$ from 1 to 2. An $n$=2 fit captures the band average expected foreground brightness temperature ($a_0$), a power law spectral shape ($a_1$), and a self-absorption correction term ($a_2$). Adding additional $a_k$ terms has minimal impact on the overall residual levels and we use daily $n$=2 fits for the rest of the analysis. 

\begin{figure}[!th]
\flushleft
\scalebox{0.43}{\includegraphics{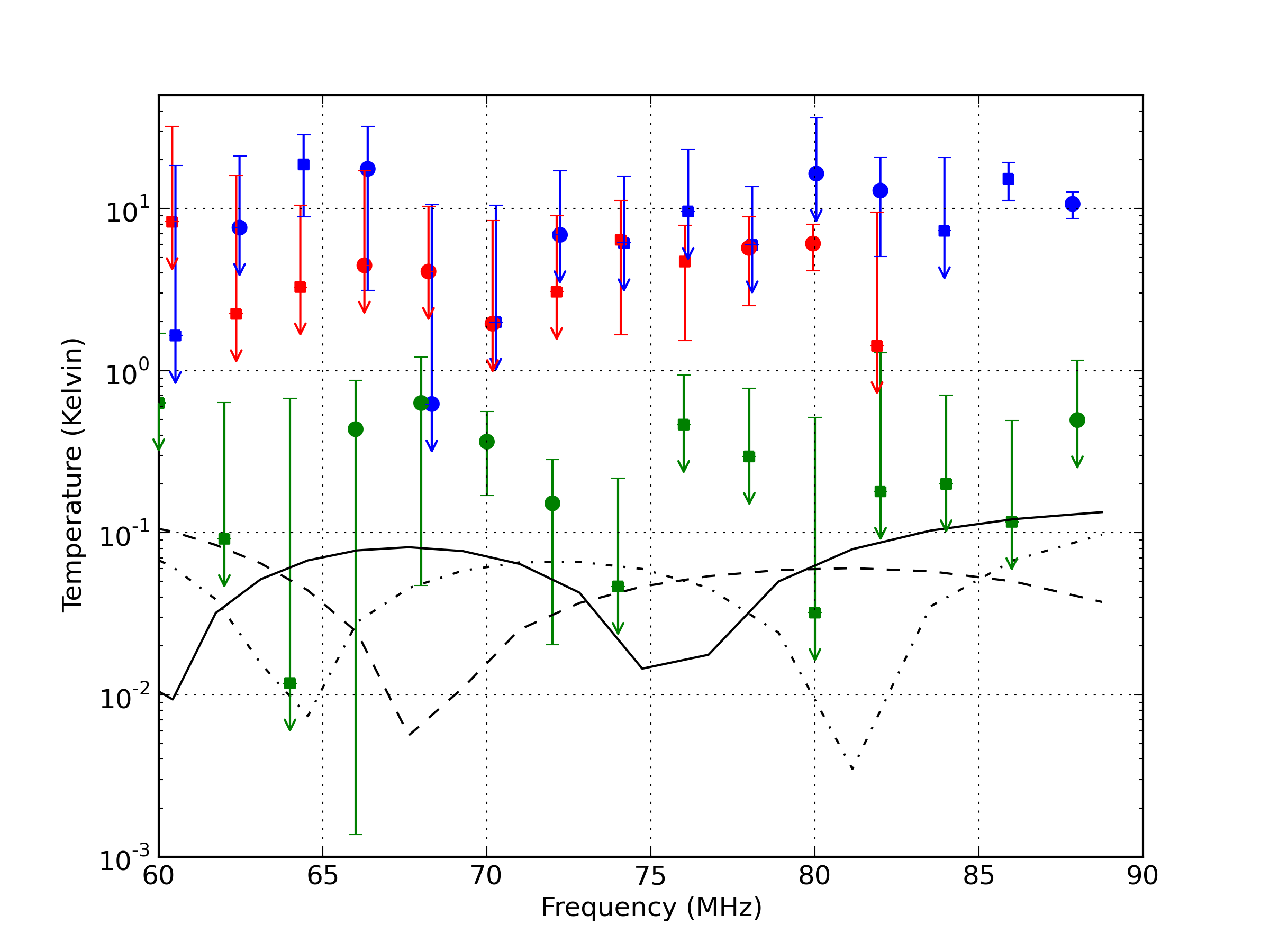}}
\caption{The log of the magnitude of the combined residuals from 4.4 hours of integration time using three different calibration techiques. Green is the residuals from $K_{\rm GSM}$, while red is the residuals from $K_{\rm \Delta GSM}$ and blue is the residuals from $K_{\rm JNC}$. Circles are positive values and squares are negative values. $K_{\rm \Delta GSM}$ fits are done on a smaller frequency range of 60-82 MHz. Error bars show the daily variance of the residuals. Also shown are the predictions from three reionization models. These models differ in their star formation efficiency and X-ray heating and the mean brightness temperature is subtracted from each of the theoretical models. \label{comb_resid}}
\end{figure}

Fig. \ref{comb_resid} shows the log of the magnitude of the mean and variance, weighted by exposure time, for fits performed with $n$=2 and $K_{\rm JNC}$, $K_{\rm GSM}$ and $K_{\rm \Delta GSM}$ calibrated data. Residuals using $K_{\rm GSM}$ are $\lesssim$1 Kelvin, particularly near $\nu$=70 MHz, while residuals using $K_{\rm \Delta GSM}$ and $K_{\rm JNC}$ are $\sim$10 Kelvin. Also shown are the predictions of three cosmological models, obtained using the {\scriptsize SIMFAST} code \citep{simfast} (see also \citet{21cmfast}). The mean over the frequency range 60-90 MHz is subtracted from each model spectrum since for the measured brightness temperature the mean is removed by the fit. 

The calibration and foreground fitting procedure can attenuate the \cm signal along with the foregrounds. To measure this attenuation, we used a simulation calibration technique \citep{paciga}. We added simulated \cm all-sky signals of varying magnitude to the raw data, then ran the calibration and foreground fitting process on the combined data. By comparing the residuals before and after addition of the simulated signal, we can quantify the amount of \cm signal that is removed in the calibration and foregound fitting. The $\chi^2$ fit for $K_{\rm GSM}$ is particularly pernicious at removing \cm spectral structure from the spectra ($\sim$1\% of an added \cm signal remains in the residuals), making it less than optimal for extracting \cm information. On the other hand, the $\chi^2$ fit for $K_{\rm \Delta GSM}$ causes minimal loss of \cm spectral structure ($\sim$75\% of an added \cm signal remains in the residuals). The absence of a fitting process in the $K_{\rm JNC}$ calibration allows full retention of the \cm structure, but the larger residuals in the data make it sub-optimal compared to the $K_{\rm \Delta GSM}$ technique. Therefore, the $K_{\rm \Delta GSM}$ technique is our current best calibration option.

\section{Discussion and Future Work}
Calibration followed by an $n$=2 fit to the all-sky Galactic spectrum gives residuals of no more than 20 Kelvin from 4.4 hours of integration regardless of calibration strategy. Given a mean foreground signal that is about 2000 Kelvin at 70 MHz, residuals are $<$1\% of the foreground signal. The absence of the additional spectral structure described in \citet{vedantham} from 60-90 MHz matches the Total Electron Content (TEC) data\footnote{Data found on the National Geophysical Data Center website (http://www.ngdc.noaa.gov/stp/IONO/USTEC/)} for Isla Guadalupe, which is $\lesssim$50 TEC Units for the period of observation. Expansion of the analysis to frequencies below 60 MHz will require consideration of the ionospheric ducting structure they describe. The residuals are also lower than those predicted by \citet{Liu_etal}, particularly at lower frequencies. This implies that the the $\sim$10\% variance of the GSM angular structure assumed in \citet{Liu_etal} contains significant fine-scale mapping errors such as striping due to reciever gain drift. 


Measured residuals are dependent on calibration scheme. Calibration using an external noise source ($K_{\rm JNC}$), similar to the \citet{edges} experiment technique, yields the largest residuals. This technique also has the greatest sensitivity to errors in the measurement of the transmission efficiency $\eta (\nu)$, which contribute to the residual signal. Conversely, calibration using the GSM model is relatively insensitive to errors in $\eta (\nu)$; but is more sensitive to differences between the simulated antenna beam and the actual antenna beam. 

Beam errors contribute more to $K_{\rm \Delta GSM} (\nu)$ than $K_{\rm GSM} (\nu)$. For full days of data the $K_{\rm \Delta GSM}$ technique essentially compares the peak of the Galactic brightness at LST=18hrs with the minium at LST=11hrs. However, because of limited battery life, most of the days in our dataset have only 10-15 hours of coverage. In many cases we miss either the peak or the minimum. Then the $K_{\rm \Delta GSM}$ fits become more sensitive to the antenna beam shape. This is a key reason residuals are larger in the $K_{\rm \Delta GSM}$ analysis than the $K_{\rm GSM}$ analysis.

In order to achieve measurement of the \cm signal (300 mK peak to peak temperature amplitude), it is necessary to decrease our residuals a further 1-2 orders of magnitude. Using the radiometer equation, we can see that thermal noise level is $\sim$2 orders of magnitude below the current residual levels, and also below the predicted \cm signal level. Therefore, the current residuals are dominated by systematic errors and residual foregrounds. A few experiment improvements are planned prior to future data collection to decrease the systematic errors. 

First, detailed examination of our datasets has revealed that our Faraday Cage is not performing as well as hoped at preventing self-generated RFI above 100 MHz. Some of this self-generated RFI can also be seen below 90 MHz for single days of data. This RFI can be found at different frequencies on each day, and is also measured in the system noise ($P_{\rm short}$) on that day. Improvements are therefore in progress to the sampling system and its shielding, requiring repeated deployment to quiet sites to implement. Deployment to quiet sites is necessary because the low level of the self-generated RFI makes it undetectable in lab tests, where the noise is dominated by external RFI. 

Second, as mentioned above, some of the residuals can be traced to calibration error associated with the lack of full 24 hour days of data. Improvements to the sampling system and the power generation system will facilitate the collection of complete days of data, reducing the calibration errors. 


As addressing these issues requires further deployment of the experiment; deployment is planned for Isla Socorro (Latitude $18^{\circ}$ $48'$ $0''$ N, Longitude $110^{\circ}$ $90'$ $0''$ W) or Isla Clari\'{o}n (Latitude $18^{\circ}$ $22'$ $0''$ N, Longitude $114^{\circ}$ $44'$ $0''$ W), $\sim$600 and $\sim$700 km off the Pacific coast of Mexico respectively. Further from the mainland than Isla Guadalupe, these islands are expected to provide an RFI environment where the FM band signals are below the thermal noise level. With two full weeks of data providing over 25 hours of integration, our goal is to achieve residuals below the 100 mK level across the frequency range 40-130 MHz.
 
\acknowledgments{Travel to Isla Guadalupe would not have been possible without support from local agencies in Mexico, including Grupo de Ecolog\'{i}a y Conservaci\'{o}n de Islas A.C. (GECI), Secretar\'{i}a de Marina (SEMAR), Secretar\'{i}a de Gobernaci\'{o}n (SEGOB), Comisi\'{o}n Nacional de Areas Naturales Protegidas (CONANP), Reserva de la Biosfera de la Isla Guadalupe, Sociedad Cooperativa de Producci\'{o}n Pesquera de Participaci\'{o}n Estatal Abuloneros y Langosteros, S.C.L., and Dr. Ra\'{u}l Michel. 

A.N., J.B.P., and T.C.V. acknowledge funding from NSF grant AST-1009615. A.N. thanks the McWilliams Center for Cosmology for partial financial support. J.M.G.C. and O.L.-C. thank INAOE for financial support. J.M.G.C. acknowleges a CONACyT Beca Mixta that allowed him to spend a year at the McWilliams Center for Cosmology}

\bibliographystyle{apj}
\bibliography{references}

\end{document}